\documentclass{article}
\usepackage{breckenr2005}
\usepackage{graphicx}
\usepackage{amssymb}
\frompage{000} \topage{000}
\def\rightmark{Resonance Production in RHIC Collisions}
\def\leftmark{C. Markert}

\title{Resonance Production in RHIC Collisions}
\authors{
{C. Markert for the STAR collaboration $^1$}\\[2.812mm]
{\normalsize \hspace*{-8pt}$^1$  Physics Department, Kent State
University, Kent, OH 44242, USA  \\[0.2ex]
}}

\abstract{Results of resonance particle production measured at RHIC
in $\sqrt{s_{\rm NN}} = $ 200 GeV Au+Au collisions are compared
to measurements in p+p and d+Au collisions in order to verify the
existence of an extended hardronically interacting medium. Yield and
momentum distributions of resonances maybe modified during the fireball
lifetime due to resonance decay and the subsequent rescattering of
their decay daughters as well as the regeneration of resonances from
their decay products. Modified momentum spectra in heavy ion
collisions may change the nuclear modification factor R$_{\rm AA}$.
The influence on the elliptic flow v$_{2}$ due to late regeneration
of resonances is discussed.}

\keyword{resonance, lifetime, strange, freeze-out, medium}
\PACS{25.75.Dw}

\begin{document}

\maketitle
\setcounter{page}{1}

\section{Introduction}
\label{intro}

If the life time of the hadronic medium created in a heavy ion
reaction is long enough (on the order of a few fm/c) resonances
decay inside the medium and their decay product can therefore interact
with the medium as well. In a dense hadronic medium resonances can
also be regenerated from their decay products. This is a dynamical
process of creating and decaying resonances (detailed balance).
The measured resonances are therefore a composition of early and
late produced and decayed resonances. The largest fraction comes from
the late decay since the probability of another interaction of the
decay particles from early decays is larger than from late decays.
Therefore by using the yield of resonances and knowing their decay,
re-scattering and regeneration cross section we are able to estimate
the lifetime of the hadronic phase.

A resonance signal loss in the low momentum region is caused by a
larger re-scattering than regeneration cross section. This will also
influence the low momentum region of the nuclear modification
factor R$_{AA}$ spectrum, which is the ratio of the heavy ion
transverse momentum spectrum divided by the p+p spectrum scaled
by the number of binary collisions.

More than 60\% of the stable particles originate from weak and
resonance decays. This raises the question how the elliptic flow
v$_{2}$ of stable particles is affected if regeneration of
resonances after chemical freeze-out is present. Furthermore
recombination models predict an enhancement of v$_{2}$ for
resonances due to the larger v$_{2}$ contribution from regenerated
resonances. The yield would scale according to constituent quark
scaling of the recombining hadrons, e.g. K $+ \pi \rightarrow
$K(892) (2+2=4) and $\Lambda+ \pi \rightarrow \Sigma(1385)$
(3+2=5). The observed enhancement of v$_{2}$ would depend on the
fraction of regenerated resonances \cite{nonaka04}.

\section{Rescattering and Regeneration Cross Sections}

\begin{figure}[h!]
 \centering
\includegraphics[width=0.70\textwidth]{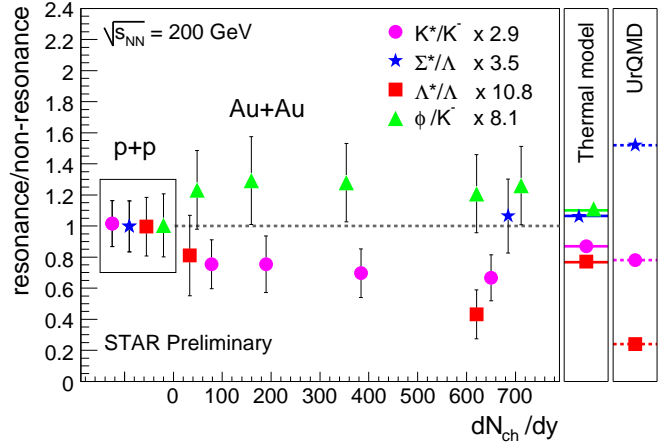}
 \caption{Resonance to stable particle ratios of $\Sigma(1385)/\Lambda$,
$\phi /K^{-}$, $\rm K(892)/K^{-}$ and $\Lambda(1520)/\Lambda$ for
$p+p$ and $Au+Au$ collisions at $\sqrt{s_{NN}} = 200\;\rm GeV$.
The ratios are normalized to unity in $p+p$ collisions. The
quadratic sum of statistical and systematic uncertainties are
included in the error bars. Thermal and UrQMD model predictions
are presented as well \cite{bec02,ble02,ble03}.
} \label{ratio1}
\end{figure}

Figure~\ref{ratio1} shows the resonance to stable particle ratios
for p+p and Au+Au collisions systems ~\cite{resostar,sevilphd}.
The ratios are normalized to unity for $p+p$ collisions. The
deviation of the ratio from unity in Au+Au collisions indicate a
late hadronically interacting medium where decay of resonances and
the rescattering of the decay particles is larger than the
regeneration of resonances. A ranking of the overall regeneration
over rescattering cross section can be deduced as follows: $R
_{\Lambda(1520)} < R_{K(892)} < R_{\Sigma(1358)}$.

\begin{figure}[h]
 \centering
\includegraphics[width=0.70\textwidth]{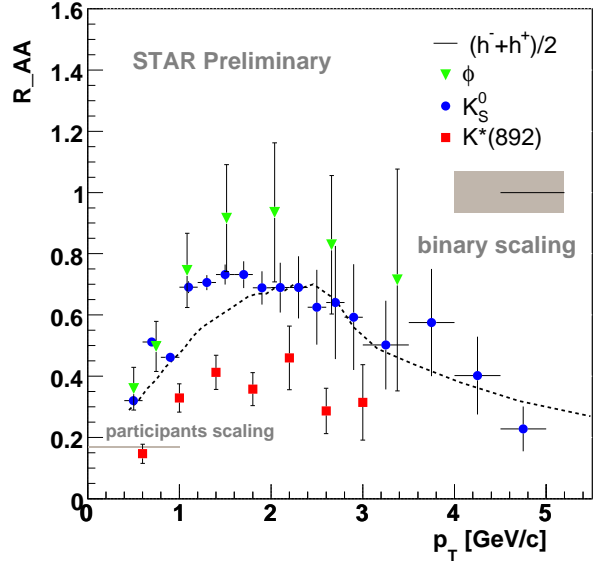}
 \caption{The ratio of the transverse
momentum of Au+Au divided by p+p collisions normalized to the
number of binary collisions.} \label{raa}
\end{figure}

Sascha Vogel showed at this workshop a microscopic model
calculation (UrQMD) of the regeneration cross sections of
resonances and confirms the ranking as derived from the data
\cite{sascha06}. This model predicts a signal loss in the low
momentum region due to rescattering. A comparison of the
transverse momentum spectrum of p+p and Au+Au collisions using the
nuclear modification factor R$_{\rm AA}$ shows for the K(892) a
larger suppression in the momentum region from p$_{\rm T}$ =
0-2~GeV compared to the long lived $\phi$ and the K$^{0}_{\rm S}$
(Figure~\ref{raa}). These data support the concept of a hadronic
interacting medium after chemical and before kinetic freeze-out
which changes the measured resonance yield and the momentum
spectra. Thermal model predictions by W. Florkowski et. al show
a shift to higher values compared to the measured K(892)
transverse momentum distribution in the low momentum
region \cite{flo04}. Therefore nuclear suppression factors
(R$_{AA}$) of resonances are not directly comparable to stable
particles as long as the momentum dependent signal loss of resonances
in the hadronic phase is not taken into account.

\begin{figure}[h!]
 \centering
\includegraphics[width=0.70\textwidth]{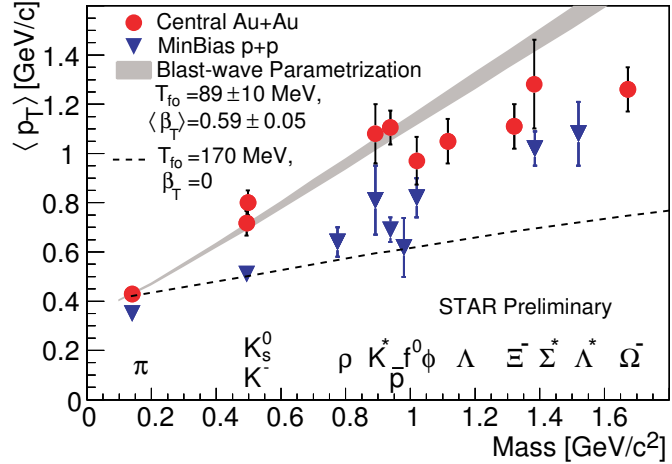}
 \caption{The $\langle p_{T} \rangle$  vs
particle mass measured in min-bias $p+p$ and most central $Au+Au$
collisions at $\sqrt{s_{NN}}=$ 200 GeV. Statistical and systematic
uncertainties are shown \cite{resostar,sevilphd}.} \label{meanpt}
\end{figure}

The low mean transverse momentum for heavy multi strange particles
($\Xi$ and $\Omega$) in Au+Au collisions was interpreted as a sign
of early decoupling from the hadronic system at high temperature
and low transverse velocity. In Figure~\ref{meanpt} the
$\Sigma(1385)$ shows the same trend as $\Xi$ and $\Omega$. Since
the regeneration cross section of the $\Sigma(1385)$ from
$\pi$+$\Lambda$ is very large \cite{sascha06} we expect to measure
more late produced $\Sigma(1385)$, i.e. resonances produced close
to the kinetic freeze-out. This would suggest a late decoupling of
the $\Sigma(1385)$ from the hadronic phase, and thus possibly
contradict the simple connection between $<$$p_T$$>$ and the
decoupling parameters. Based on the primordial $\Lambda$ spectra,
a $\Lambda$ from a $\Sigma(1385)$ decay is not expected to exhibit
much early decoupling. Thus it is important to measure the
$\Xi(1530)$ resonance in heavy ion collisions, verify its
regeneration cross section and estimate the contribution of
$\Xi$'s from the $\Xi(1530)$ decays.

\section{d+Au Collisions}

\begin{figure}[h!]
 \centering
\includegraphics[width=0.49\textwidth]{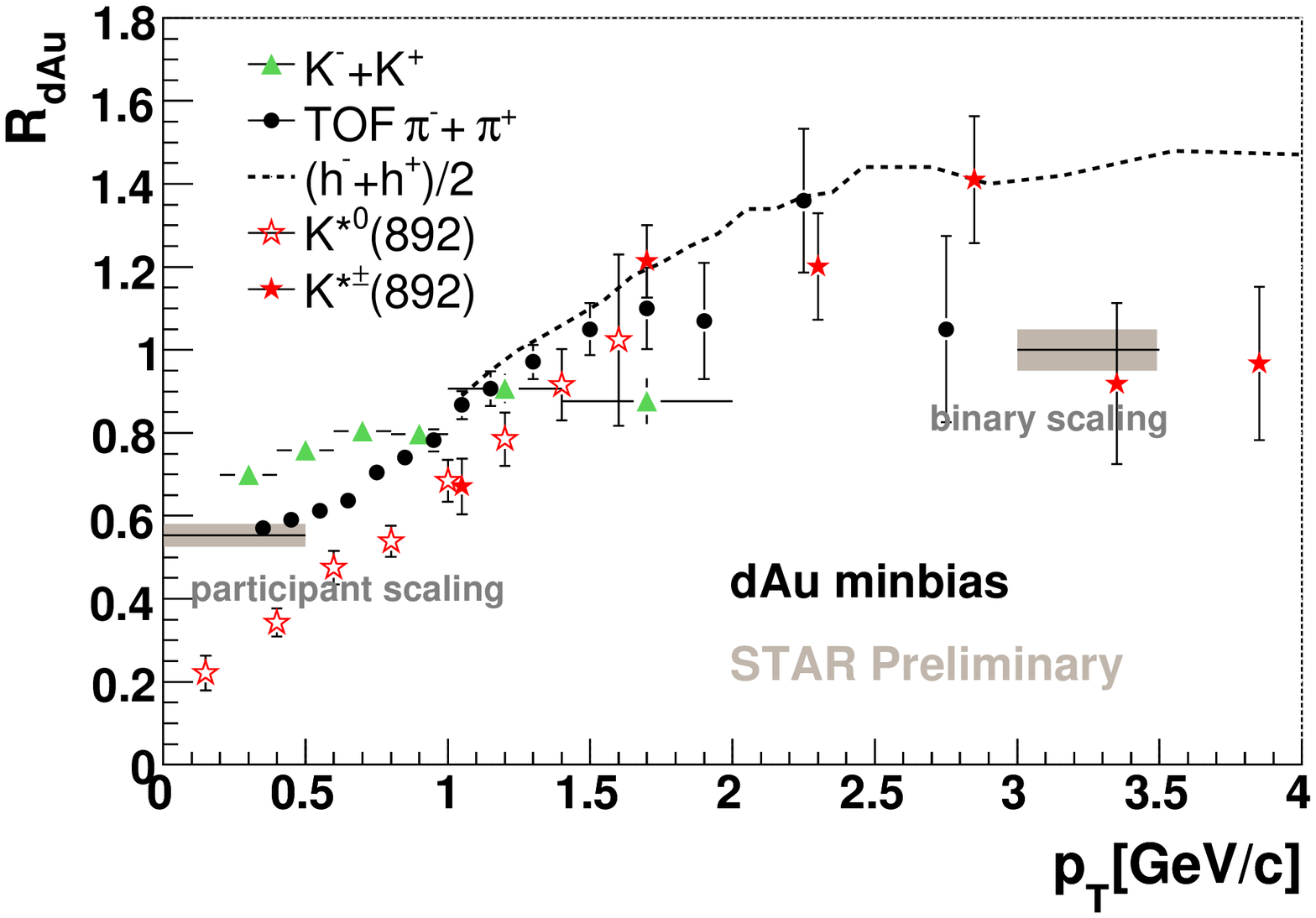}
\includegraphics[width=0.49\textwidth]{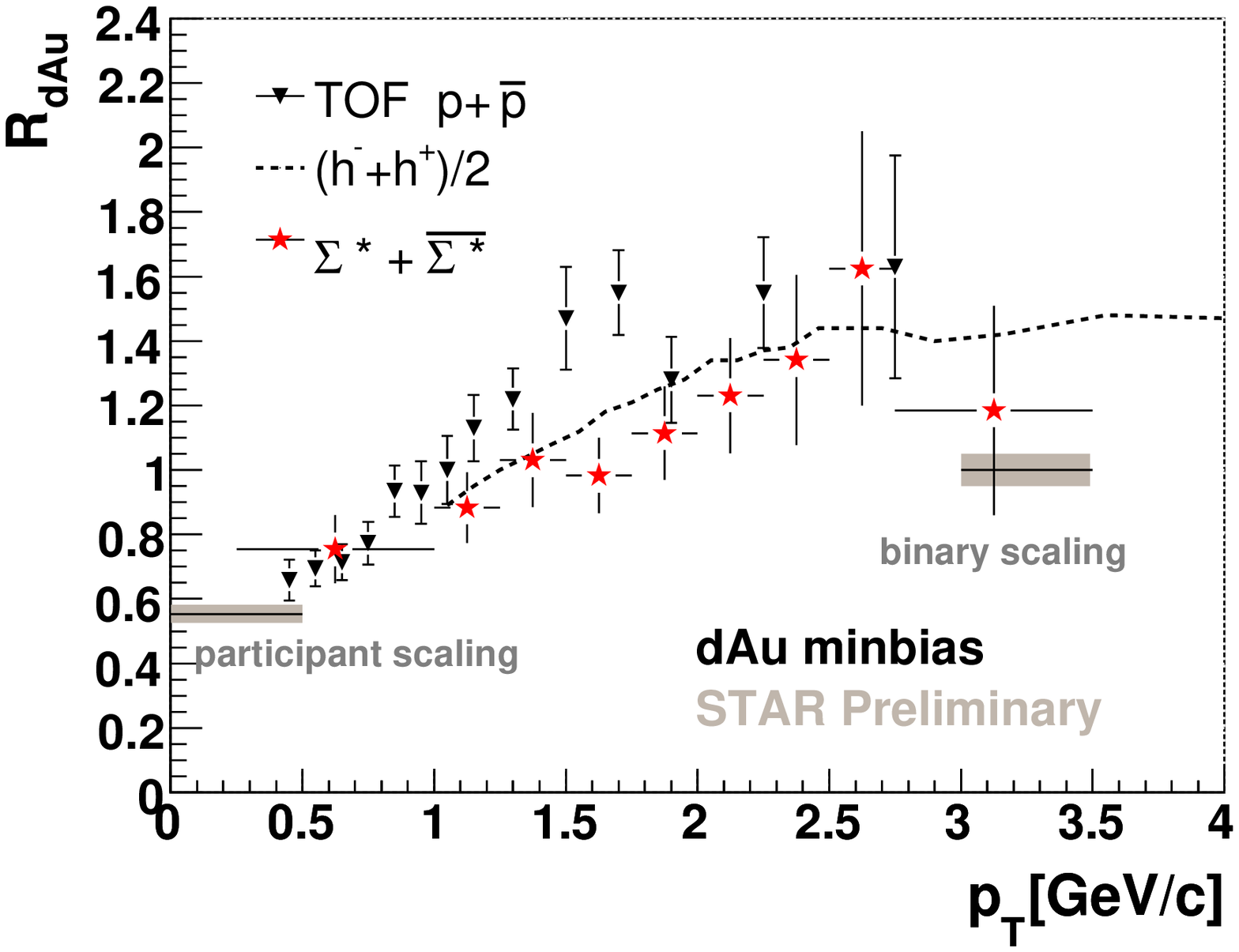}
 \caption{The ratio of the transverse
momentum of d+Au divided by p+p collisions normalized to the
number of binary collisions for meson (left) and baryons (right).}
\label{rdau}
\end{figure}

Since no extended hadronic medium for d+Au collisions is created,
the momentum distribution of p+p and d+Au collisions for
resonances are expected to be similar compared to the stable
particles. Figures~\ref{rdau} shows the preliminary ratio of the
transverse momentum spectra measured in d+Au divided by
p+p spectra normalized to the number of binary collisions.
The R$_{dAu}$ of $\Sigma(1385)$ seams
to follow the proton R$_{dAu}$ (right), while the K(892) deviates
from the K and $\pi$ in the low momentum region. This result is
unexpected, if we assume no rescattering in d+Au. The data
have to be further investigated and the yields have to be checked
for consistency.

\section{Elliptic Flow v$_{2}$}

Anisotropic flow results are often used as a strong evidence for
the formation of the QGP in Au+Au Collisions at RHIC. The
magnitude and centrality dependence of the elliptic flow v$_{2}$
is used as a proof of early thermalization. The so-called "mass
splitting", the characteristic dependence of v$_{2}$(p$_{\rm T}$)
on the particle mass, is well described when using a QGP Equation of
State. In addition the constituent quark scaling in the intermediate
transverse momentum region is often cited as a proof of
deconfinement and partonic (pre-hadronic) collectivity. One therefore
would like to study the elliptic flow of resonances in order to
test the probability of formation at the early stage of the collision
and the effect the expansion dynamics of the source including
rescattering and regeneration of resonances will have on their v$_2$.
The elliptic flow of the long lived $\phi$ meson resonance shows the
expected mass dependence in the low momentum region and a
constituent quark scaling similar to the K$^{0}_{S}$ meson in the
intermediate transverse momentum region. In the near future we will
analyze the elliptic flow of the short lived K(892) and $\Delta(1232)$
resonances.

\begin{figure}[h!]
 \centering
\includegraphics[width=0.55\textwidth]{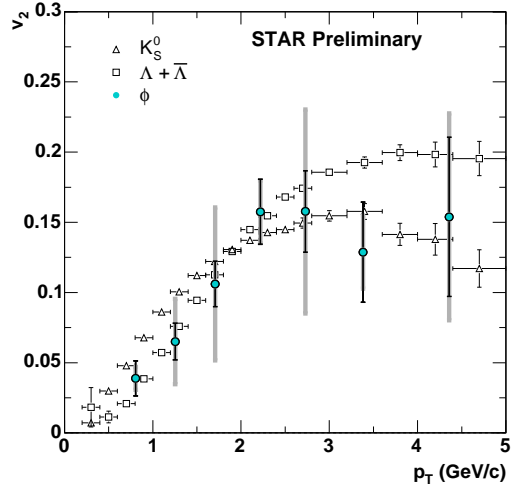}
 \caption{The v$_{2}$ of $\phi$ meson compared to K$^{0}_{S}$ meson and
 $\Lambda$ baryon  as a function of p$_{\rm T}$. Statistical and
systematical
 errors are included \cite{phiqm2005}. }
\label{phiv2}
\end{figure}

\section{Conclusions}

Resonances are a unique tool to probe the hadronic medium in heavy
ion collisions. It allows us to estimate the lifetime of the
hadronically interacting medium and derive the rescattering over
regeneration cross section ranking for different resonances.
Using a microscopic models actual regeneration cross sections can
be determined based on the data. Further more we gain in a better
understanding of the hadronic interaction probabilities which helps
us to distinguish between early and late decoupling of particle
species from the hadronic medium.

\section*{Acknowledgments}
I would like to thank Sascha Vogel and Marcus Bleicher for their
detailed UrQMD analysis to get a better understanding of hadronic
cross sections in terms of resonance regeneration. And I also
would like to thank the STAR collaboration for the support in
presenting this data.

\vfill\eject

\begin{thebibliography}{99}

\bibitem{nonaka04} C. Nonaka et. al., Phys.Rev. {\bf C69} (2004) 031902,
nucl-th/0312081.
\bibitem{resostar}  J. Adams, {\it et al.}, (STAR collboration), Subm. to Phys.
Rev. Lett, nucl-ex/0604019.

\bibitem{sevilphd} Sevil Salur, PhD Thesis Yale University 2006.

\bibitem{bec02} F. Becattini, Nucl. Phys. {\bf A702}, 336 (2002).
\bibitem{ble02} M. Bleicher {\it et al.},  Phys. Lett. {\bf  B530}, 81
(2002).
\bibitem{ble03} M. Bleicher and H. St\"ocker, {\it J. Phys.} {\bf G30}
(2004) 111.
\bibitem{sascha06} S. Vogel, These proceedings.
\bibitem{flo04} W. Florkowski, W. Broniowski and P. Bozek, J.Phys. {\bf
G30} (2004) 1321.

\bibitem{salur06} S.Salur (for the STAR Collaboration)
nucl-ex/0606002.
\bibitem{phiqm2005} X. Cai for the STAR collaboration, Quark Matter 2005,
Budapest, Hungary, 4-9 Aug 2005. e-Print Archive:
nucl-ex/0511004



\end{thebibliography}
\end{document}